\documentclass[11pt]{article}
\usepackage[a4paper,text={5.8in,9.7in},centering]{geometry}

\usepackage{epsf,epsfig,graphicx}
\usepackage{makeidx} 

\usepackage{wrapfig}

\usepackage{authorindex}

\usepackage[applemac]{inputenc}

\usepackage{array}
\usepackage{units}
\usepackage{textcomp}
\usepackage{amsmath}
\usepackage{amssymb}
\usepackage{esint}
\usepackage{amsfonts}
\usepackage{multirow}
\usepackage{rotating}

\usepackage{subcaption}

\usepackage{wrapfig}
\usepackage{sidecap}
\usepackage{color}
\usepackage{caption}
\usepackage{paralist}
\usepackage{verse}

\usepackage{endnotes}
\makeatletter
\renewcommand*\makeenmark{\hbox{\textsuperscript{\@alph{\theenmark}}}}
\makeatother

\date{}

\newcommand{\be}{\begin{equation}}
\newcommand{\ee}{\end{equation}}
\newcommand{\ben}{\begin{displaymath}}
\newcommand{\een}{\end{displaymath}}

\begin{document}

\title{\bf{Oral History Interview with Marcello Cresti}}

\author{\em by Alessandro De Angelis, August 2019}

\maketitle

\vskip 0.5 cm

{\em 
Born in Grosseto, 26 April 1928, Marcello Cresti graduated in Physics in 1950 at the University of Pisa by attending the Scuola Normale Superiore, and became professor of Experimental Physics in Padua in 1965.

His research focused on the physics of cosmic rays and elementary particles.  In 1951, working at the mountain observatory built by the Padua research group at Pian di Fedaia, close to the Marmolada mountain, at an altitude of 2000 m a.s.l., he developed and used a system of overlapping Wilson chambers that allowed an analysis of particles produced by cosmic rays. With this device, in collaboration with Arturo Loria and Guido Zago, he obtained results on the extended air showers of cosmic radiation and on the production of pions and strange particles.\footnote{Cresti M., Loria A. and Zago G., ÒCamera di Wilson in campo magneticoÓ, Nuovo Cimento 10, 843 (1953); Cresti M., Loria A. and Zago G., ``Sulla distribuzione zenitale degli sciami estesiÓ, Nuovo Cimento 10,  779 (1953).}  As other mountain observatories, the Fedaia Laboratory had an important role in establishing international collaborations with other cosmic ray groups in Europe. In 1955 he worked at the Max Planck Institute for Physics in Gšttingen, directed by Werner Heisenberg, to analyze the data obtained, developing a technique for reconstructing events that used one of the first electronic computers, built in that Institute by Heinz Billing in the early 1950s. During his stay in Gšttingen, he collaborated with Martin Deutschmann who brought to the Fedaia observatory a multiplate cloud chamber\footnote{Deutschmann M., Cresti M., Greening W.B., Guerriero L., Loria A. and Zago G.,``An Anomalous V$^0$ EventÓ, Nuovo Cimento 3,  180 (1956).}  and also established a relationship with Klaus Gottstein, who later led the Max Planck cosmic ray group and with whom he later collaborated in Berkeley.   

In 1956/57 he worked at the UCRL (University of California Radiation Laboratory) in Berkeley under the direction of Luis W. Alvarez, and carried out the first experiments with liquid hydrogen bubble chambers. In this group he brought his experience in the use of Wilson's chambers and in the automatic analysis of the photos obtained. Alvarez's group devised and used for the first time in a systematic way automatic measurement, which were made possible by the progress made at that time by electronic calculators. The use of these techniques, and of the hydrogen bubble chamber, allowed the completion of an important experiment on the production of strange particles and its rapid and complete analysis, which led to many results of fundamental importance, in particular the discovery of the first interaction violating parity and not involving neutrinos: the asymmetry of the decay of the Lambda baryon.\footnote{Crawford F. S., Cresti M., Good M.L., Gottstein K., Lyman E.K., Solmitz F.T., Stevenson M.L., Ticho H.K., ``Detection of parity nonconservation in $\Lambda$ decayÓ, Physical Review, 108,  1102-1103 (1957).}  During his second year in Berkeley he was staff member.

In 1958 he returned in Padua, and he set up a laboratory for the design and analysis of data taken in experiences with accelerators, and started a collaboration with CERN in Geneva; he built the first European instrument for the automatic measurement of bubble chamber photographies.\footnote{Bettini A., Cresti M. et al., ÒDescription and Operational Performance of the Padova PEPRÓ Proc. International Conference on Computer Assisted Scanning ConferenceÓ, Padova 1976, and references therein.}  

His group also built the first mass electrostatic separator in Europe, with which pure and rather monochromatic beams of antiprotons could be obtained. After 1965 he began a series of measurements on the interaction of antiprotons on deuterium.\footnote{Bettini A., Cresti M. et al., ÒThe Annihilation $\bar p n \rightarrow   \pi^+ \pi^- \pi^-$   between 1.0 and 1.6 GeV/$c$ and its comparison with Veneziano modelÓ, Nuovo Cimento 1A, 334 (1971).}

He committed himself to have an appropriately sized calculator available in Italy; together with Giuseppe Mannino, director of the astronomical observatory of the University of Bologna, he was the architect of the Interuniversity Consortium for the Automatic Computing of Casalecchio, Bologna.  

In 1975-76 he was at CERN in Geneva, where he designed and built a low-energy antiproton beam line with excellent monochromaticity and collimation to be used for high statistics and resolution measurements involving antiprotons.\footnote{Cresti M., Pasquali G., Peruzzo L., Pinori C., and Sartori G., ``Measurement of the Antineutron MassÓ, Physics Letters 177B, 206 (1986).}   

From 1976 to 1980 he contributed to the creation of the European Hybrid Spectrometer at CERN, a detector run by an international collaboration. The detector included a bubble chamber and a spectrometer consisting of Cherenkov, drift chambers, ionization meters, three calorimeters (one built in Padua) and TRDs, mostly used for studies of the charm quark.\footnote{Allison W.W.M. et al., proposal CERN/SPSC/76-43 (Geneva 1976); Aguilar-Benitez M. et al., ``Proposal to study the hadronic production and the properties of new particles with a lifetime $10^{-13}$s $<\tau<10^{-10}$s using LEBC-EHSÓ, CERN-SPSC-79-80 (Geneva 1979).}  

From 1981 to 1984 he was Dean of the Faculty of Sciences and from 1984 to 1987 he was Rector of the University of Padua.

From 1987 he collaborated with the group of Ugo Amaldi at CERN in Geneva for the DELPHI experiment at the LEP accelerator; the Padua group was responsible of the construction of one of the electromagnetic calorimeters.  

Since 1989 with his group in Padua and a group from Pisa he designed and carried out an experiment aimed at the detection of high energy cosmic rays, gamma rays in particular; the experiment, called CLUE, started the Italian astrophysical activity at the Roque de los Muchachos observatory in the Canary Island of La Palma.\footnote{Cresti M. et al., ÒVHE cosmic ray spectroscopy by detection of Ultraviolet Cherenkov Light (CLUE proposal)Ó, INFN Note (October 1990).}  

He retired from University in the year 2000.
}

\vspace{15pt}
\hrule
\vspace{6pt}

\vskip 3mm

{\em Alessandro De Angelis: Can you shortly summarize your pre-University time (until 1946)? Did you have any special interest in science or technology/mechanics as a child or a teenager? Why did you choose physics?}

\vskip 1 mm Marcello Cresti: As a high school student I was quick in learning mathematics and physics and was attracted by technical things. I don't remember being influenced by teachers. I wanted to become an engineer, but the cost of going to the University  would have been too high for my family, as Grosseto, my home town, didn't have one. So I applied for a scholarship at the Scuola Normale Superiore in Pisa that would have allowed me to study without any financial burden for my family. As the Normale did not have any available student position for engineering, I decided to enrol in mathematics and physics, knowing that the curricula of the first two years were almost identical for the two degrees, and change to engineering later.
After I started, I discovered that there was a course in ÒFisica PuraÓ and changed to that. I soon discovered that physics was what I really liked.

\newpage {\em ADA: Actually you enrolled in the late 1940s, when the period of post-war reconstruction of science in Europe was starting. How did you choose between experimental and theoretical physics? What was the subject of your thesis and how did you choose it?}

\vskip 1mm MC: I was not very good at theory, so the choice of an experimental subject for my thesis was obvious. 
Pisa's physics Institute was still recovering from the war. Most of the few teachers were approaching retirement age and no significant research was being performed then. 
I was fascinated by the work that was being done at that time on cosmic rays, so I asked Professor De Donatis for a thesis on that subject. 
I began to work on a cosmic ray detector, using Geiger-MŸller counters and coincidence circuits. The circuit was built by a technician using old electronic tubes left there by the US army. We built ourselves the counters using glass tubes with insides coated with aluminum. The technique was completely  unknown in Pisa, so I had to invent a way to proceed, going to Firenze and to Livorno to find people that could build and coat the counters. 
The thing became very difficult for me, as Professor De Donatis' wife became then very ill and he had to leave me practically alone. The circuit and counters were built and the experiment started. 
The only result of the work was to get me a degree.

\vskip 3mm {\em ADA: Who was working in Pisa at the time? What kind of research was being done?}

\vskip 1mm MC: Pisa's physics Institute was still recovering from the war. No significant research was being performed then. Just about when I graduated, Marcello Conversi was appointed professor and went there, starting an experimental activity which was limited at that time. This is why in 1951 I moved to Padua.

\vskip 3mm {\em ADA: Why Padua?}

\vskip 1mm MC: I moved to Padua because Professor Rostagni, who was the successor of Bruno Rossi in Padua, was looking for young people and Professor Aldo Ghizzetti, a mathematician friend of his, who had been giving at the Normale seminars that I attended, had suggested to him my name.

\vskip 3mm {\em ADA: Can you describe your impressions in arriving at PaduaÕs Institute for Physics? Who was working there at the time? Did you have any influence on the development of the Institute?}

\vskip 1mm MC:  In 1951 the Institute of Physics in Padua, under the leadership of Professor Rostagni, was developing into one of the most important in Italy and was one of the four labs to give birth to the Italian Nuclear Physics Institute  (INFN). I was one of the youngest members and as such I did not have a particular influence in its development.

\vskip 3mm {\em ADA: What do you remember of your scientific researches in Padua during your first years there? What about your colleagues?}

\vskip 1mm MC: At Padua, after the first few months of getting adjusted, in the month of August 1951 I went to a mountain lab, on the Passo Fedaia below the Marmolada, to operate a cloud chamber in a magnetic field in collaboration with Arturo Loria, Luciano Guerriero (later to become the President of the Italian Space Agency) and Guido Zago. The setup had been started by Bruno Rossi, in particular with the construction of an electromagnet, before the fascism had forced him to leave in 1938. The magnet had been built by Giacomo Someda, a professor of Electric Engineering in Padua, and RossiÕs setup was the first in the world employing such a device. Zago, with a little help from me, finished it and made it operational. We installed inside the magnet a cloud chamber constructed by Rossi and we studied the properties of the interactions of cosmic rays in the atmosphere. 
 
\begin{figure}
\begin{center}\includegraphics[width=0.8\columnwidth]{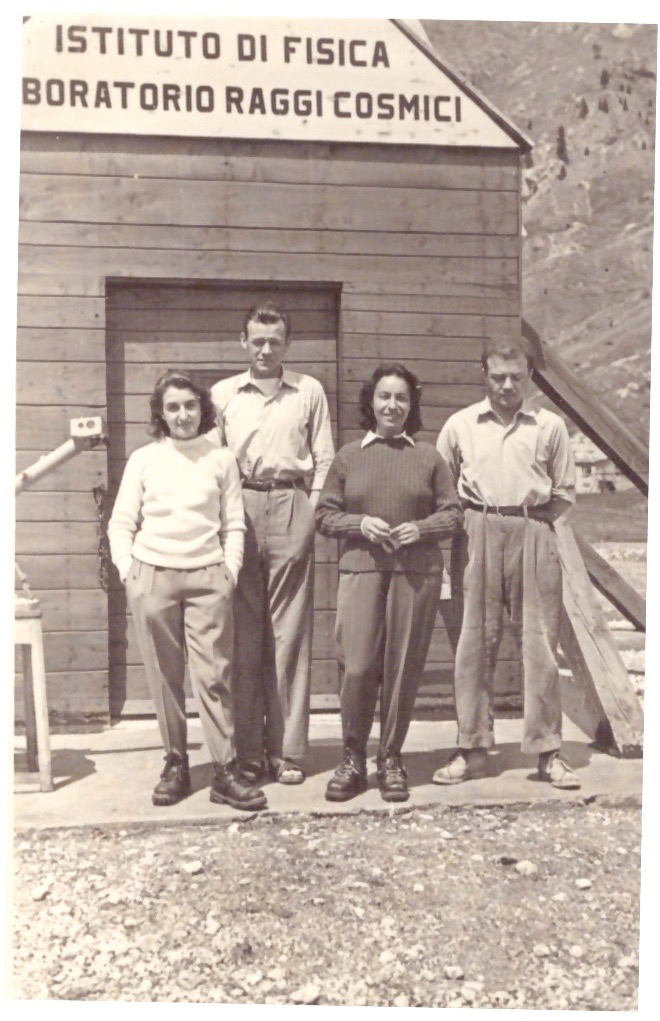}\end{center}
\caption{Cresti (left), Zago (right) and two tourists at the door of the Laboratorio Raggi Cosmici of the Padua University near Lago Fedaia, in 1952. MC: Tourists were stopping frequently at the laboratory, being curious about the meaning of the signboard.}
\end{figure}

\newpage

{\em ADA: Why did you work mostly on cosmic ray physics?}

\vskip 1mm MC: When I started there were no accelerators in Italy, so the choice was obvious. My dissertation, at Pisa, was on cosmic rays and Padua hired me to work on this subject, as that was, under Rostagni, the main topic of its research. The move to accelerators was obvious, as cosmic ray research had developed into particle (strange particles mainly) physics.

\vskip 3mm {\em ADA: As a young researcher, did you get any feeling about the ÒRossi affaireÓ in Padua and about the Rossi/Rostagni relation?}

\vskip 1mm MC: About  the ÒRossi affaireÓ  I felt the same as I did about the general nazi-fascist behavior against the jews, of which that was a small, although regrettable, part. My work in the first five years was on cosmic rays, mostly with cloud chambers, with a brief parenthesis in nuclear emulsions. The laboratory in Padua had some scientific interest also in an international framework, and was visited by many US and European scientist, in particular from the Max Planck Institute.

\begin{figure}
\begin{center}\includegraphics[width=0.84\columnwidth]{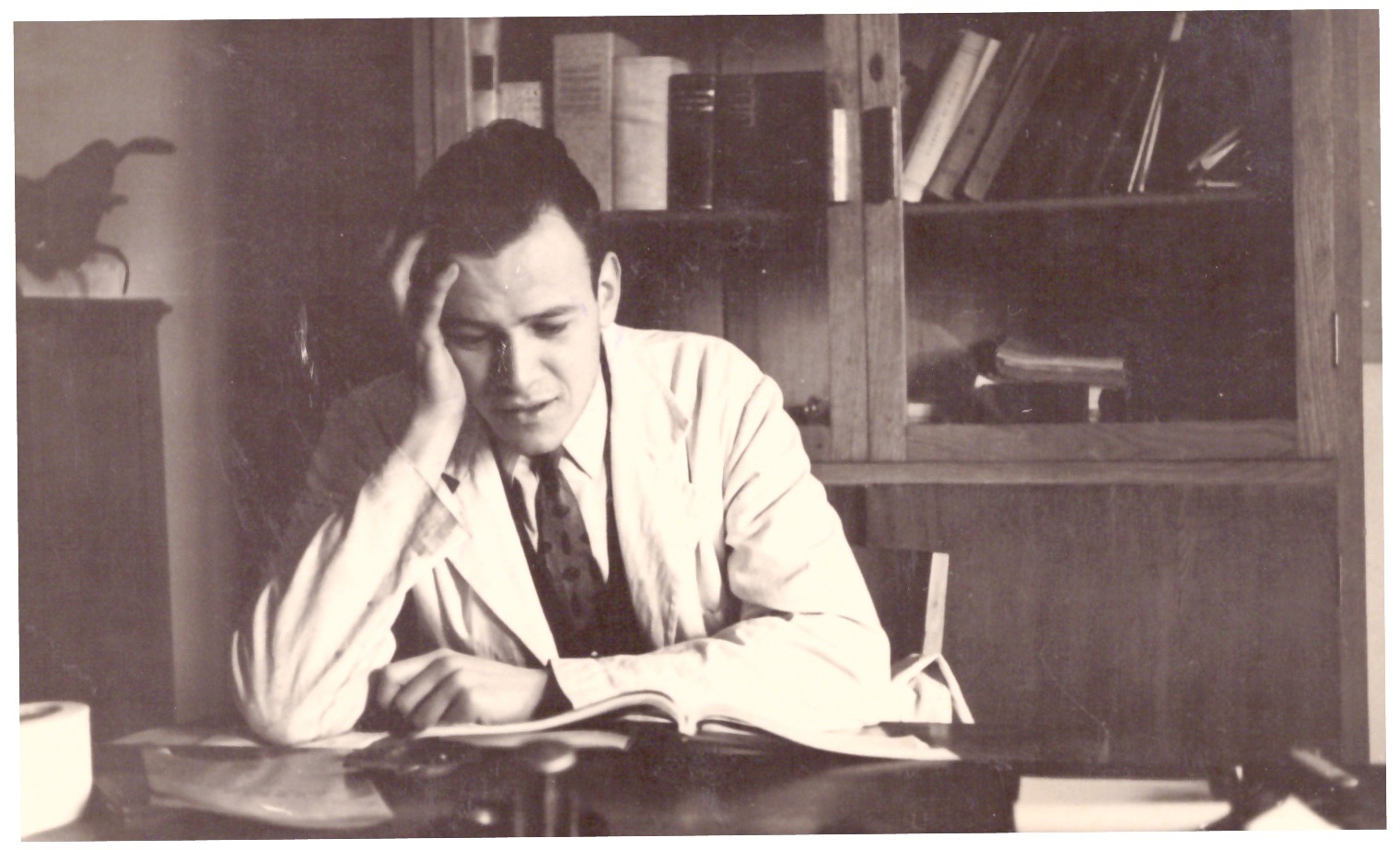}\end{center}
\caption{Cresti sitting at his desk in Padua (1953).}
\end{figure}

\vskip 3mm {\em ADA: Can you tell us about the collaboration between Padua and the Max Planck Institute for Physics in Gšttingen?} 

\vskip 1mm MC: After a couple of years of work on cosmic rays, we started a collaboration with Martin Deutschmann, a German physicist from Gšttingen's Max Planck Institute and a collaborator of Heisenberg; Deutschmann brought to Marmolada a multiplate cloud chamber that was put under our chamber inside the magnet. We collected pictures of nuclear interactions producing strange particles. 
To analyze these, I spent a few months in Gšttingen. There I found Marcello Ceccarelli, who came from Padua to work with Klaus Gottstein in the nuclear emulsion group. The experience I had gathered analyzing the V particles in the cloud chamber induced Marcello to ask me to join the group in measuring kaon-proton interactions in a stack exposed to an accelerator. I started then a collaboration and a friendship with Klaus, with whom I collaborated again later when we met at the   Radiation Lab in Berkeley.

\newpage
{\em ADA:  During the early 1950s Edoardo Amaldi and Gilberto Bernardini, together with a few others, were taking upon themselves the responsibility of the reconstruction of physics in Italy after the war. Did you have any relation  with them?}

\vskip 1mm MC: I met Edoardo Amaldi and Gilberto Bernardini, but was too young then to be friend with them. Had good relations with both of them later. Amaldi was the main motor behind the creation of INFN and Bernardini was one the outstanding figures taking part to it. Several other senior physicists, like Rostagni, were instrumental to the process.

\vskip 3mm {\em ADA: Definitely,  and within this context they also built an electron synchrotron and a National Laboratory in Frascati in 1955, and later, in 1960 (after a process started by the University of Padua in 1956), in Legnaro...
Up to the  1950s, cosmic rays, nuclear physics and particle physics were still very much related. What determined the conceptual passage to particle physics as a well-defined field apart from the existence of more powerful accelerators as new research tools providing the context for such development? How comes that you started with cosmic ray particle physics and then you made a first move to accelerator particle physics?}

\vskip 1mm MC: The transition from cosmic rays to accelerator physics was in my mind a general process due to the fact that cosmic ray research had led to the discovery of strange particles and the research on these was made easier by the accelerators. I followed this transition.

\vskip 3mm {\em ADA: And how could yourself transform from a cosmic ray physicist into a ``particle physicist''?}

\vskip 1mm MC: In the beginning at Berkeley my experience in analyzing cloud chamber pictures was useful to the group that was collecting bubble chamber pictures. Then I got back much more than I had put in, like working at an accelerator, using computers and getting knowledge in the use of sophisticated techniques, like computers, automatic measuring devices, and fast electronics, that were new to me.

\vskip 3mm {\em ADA: What are the main scientific results of that time?}

\vskip 1mm MC: The main results of the Alvarez group at UCRL in Berkeley in those years were on production and  decay of strange particles. The outstanding result was the first discovery of parity non-conservation in interactions not involving neutrinos.

\vskip 3mm {\em ADA: All this happened in the late 1950s, which marked the beginning of the space era... At that time you were in the US, when the launch of the Sputnik was announced. What do you remember of the impact of such news in the US research centers?}

\vskip 1mm MC: The reactions to Sputnik were mixed. Partly admiration and partly envy, together with rage for having been beaten...

\vskip 3mm {\em ADA: Did you know Bruno Rossi? Did you have any interaction with him when in the US?}

\vskip 1mm MC: I met Bruno Rossi only once and briefly.

\vskip 3mm {\em ADA: How can you compare the conditions of work in the US with the conditions in Padua?}

\vskip 1mm MC: The conditions were very different, mostly due to financial reasons.

\vskip 3mm {\em ADA: Why did you come back from US to Italy?}

\vskip 1mm MC: I came back to Italy for personal reasons: I wanted to return home.

\newpage {\em ADA: Can you detail a bit your contribution to research after coming back to Italy in 1958? Bubble chamber, computers, electrostatic separator, etc...}

\vskip 1mm MC: My contribution to research has been mainly on the technical side. The experience I had gathered in the US, together with the help I got from the friends I had made there, helped me to construct some instruments that turned out to be very useful. The contact with the electronic computers, at the time not very common in Italy, was very important in helping me in my first experiments with bubble chambers. I did not get any outstanding results, but some of the results my group obtained a good place in the literature of the period.

\vskip 3mm {\em ADA: What is the impact on you of the computer, and the influence on your way of thinking experiments?} 

\vskip 1mm MC: The  computer, and in general the technical development of instrumentation, did not change my way of thinking, it only made it easier to do what we were thinking of.

\vskip 3mm {\em ADA: When in Padua, what was your relation with the other researchers, Milla Baldo Ceolin in particular?}

\vskip 1mm MC: My relations with the other researchers were generally good, in some instances, Milla was one of them, they have been of a very good and long-lasting friendship. Milla helped me to get recognition in the lab and contributed to my career.

\vskip 3mm {\em ADA: How would you compare your experience at CERN in 1975-76 with the experience in the US?}
								
\vskip 1mm MC: The period you mention, almost twenty years later than the time I spent in the US, makes the comparison meaningless. The comparison is significant only if made with my first years at CERN, from 1959. In that period UCRL was a well-established laboratory run by competent people. CERN was populated by young and enthusiastic researchers. Technically were both very good; the main difference was perhaps that the US people were more professional and we Europeans were more enthusiastic.
Also, and very important, the accelerator at Berkeley was at the end of a long, very successful career, while the Proton Synchrotron  at Geneva was at the beginning of an equally successful career. It's hard, therefore, to push the comparison further.
As far as my experience goes, at Berkeley I was a beginner, whereas at CERN I had some experience, not much less than many of people working there.

\vskip 3mm {\em ADA: In the 1980s you stopped research for six years to work on University administration. Why? Do you regret it?}

\vskip 1mm MC: I did not stop research entirely, but kept on for the first three years. The reason for my involvement in University administration, as you call it, although rather amusing, has nothing to do my scientific career, and I won't recall it here.
In any case I don't regret it. It was, to mention my wife Lee's comment, a Òvery interesting observation pointÓ on the behavior of people towards us.

\begin{figure}
\begin{center}\includegraphics[width=0.84\columnwidth]{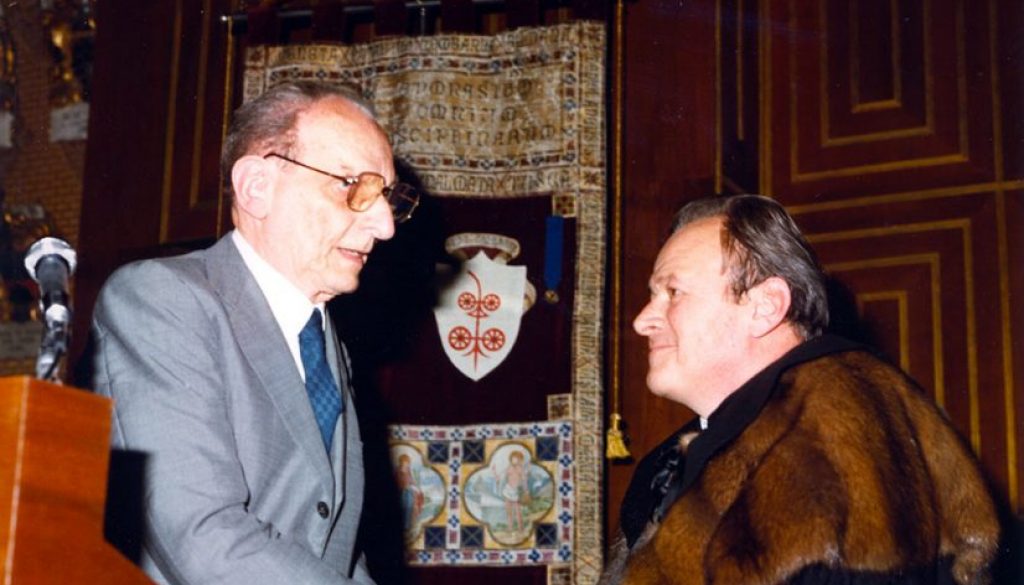}\end{center}
\caption{Cresti (right) with the philosopher Norberto Bobbio in 1985.}
\end{figure}

\vskip 3mm {\em ADA: After a successful work in administration, you first moved to a large-scale (about 400 scientists) experiment at the CERN Large Elecron-Positron collider (LEP). Why LEP? How did you live this transition?}

\vskip 1mm MC: The group I was working with in Padua joined LEP, so I stayed with them. But I did not like working in a collaboration with several hundred people...

\newpage {\em ADA: How comes that after a short secondary experience in accelerator physics you moved back to cosmic ray physics, and in particular gamma-ray astrophysics, during your last years of activity?}	

\vskip 1mm MC: As I told you I did not like working in a collaboration with several hundred people, so I took advantage of an offer to join the design and the construction of CLUE, a ground-based gamma-ray telescope to detect the ultraviolet part of the Cherenkov radiation produced by extensive showers in the atmosphere.

\vskip 3mm {\em ADA: CLUE, that you proposed with Aldo Menzione and Angelo Scribano from Pisa, used an innovative technique for the detection of gamma rays. Why this and not one of the existing ones (Imaging Air Cherenkov and Extensive Air Showers)?} 

\vskip 1mm MC: Why do people think of new experiments, rather than joining existing ones?

\vskip 3mm {\em ADA: The first site you explored for CLUE was Mauna Kea, but in the end you went to La Palma (that later became one of the two main centers worldwide for gamma ray astronomy). Why did the attempt with Hawaii fail? How comes that you ended up installing the telescope in Canary Islands?}

\vskip 1mm MC: The reason given by the Hawaii people for not accepting us on Mauna Kea was the impact of a large array of instruments on sacred ground. Canary Islands became then the most convenient substitute, as La Palma had already installations from Max Planck Institutes. The suggestion to contact the Observatory in Canary Islands came from Eckart Lorenz, who was regularly attending meetings organized in the Elba Island by the Pisa group, and was leading a strong experimental team from the Max Planck Institute ``Werner HeisenbergÕÕ in Munich (MPI). Lorenz and the Max Planck group had already deployed in 1987, on top of the Roque de los Muchachos in La Palma, an array of Cherenkov telescopes called HEGRA, also studying gamma rays but with a different technique.

\vskip 3mm {\em ADA: In the beginning a large part of your group did not follow you in this new adventure. How did you feel about this?}

\vskip 1mm MC: As far as my group, which was not my group any more, they did not share my feeling about the extremely large collaborations. They had a leading role in LEP so they continued there.

\newpage {\em ADA: The Padua group will  anyway have a leading role a few years later in the MAGIC telescopes in La Palma and now in CTA LST, still collaborating with MPI Munich. But the gamma-ray technique used in CLUE did not become the leading one in the going-to-be-leading field of gamma-ray astrophysics: imaging Cherenkov telescopes got the edge in the end. Do you regret something?}

\vskip 1mm MC: The technique we used in CLUE was based on the performance of a new UV-sensitive substance, TMAE, that we overestimated, so the results were inferior to the expectations. Of course I (and all my colleagues) weren't happy about it. I regret the failure, but enjoyed the experience.
Curiously enough the last experiment I took part in was of a nature very similar to the first one. They were both work on detecting cosmic ray showers at a high altitude, the first with a cloud chamber, the last with CLUE.

\vskip 3mm {\em ADA: Which discoveries in science did impress you most in your lifetime?}

\vskip 1mm MC: The discoveries were so many that I find it impossible to single one or even a few.

\vskip 3mm {\em ADA: Looking in perspective, what do you think have been your main personal contributions to science and when did you achieve them?}

\vskip 1mm MC: My personal contribution to science, as you call it, has been no better and no worse than that of a large fraction of the people that worked on the field. No single result was outstanding, except the one that I already mentioned, on parity non-conservation, and that can hardly be attributed to myself.

\vskip 3cm

\begin{quotation}
\noindent
{\em \footnotesize Alessandro De Angelis, graduated  in 1983 in the group of Marcello Cresti, is professor of Experimental Physics in Padua and  in 
Lisbon. He has been for several years staff member at CERN working on accelerator physics;   since   1999  he works mostly on cosmic rays, and has been project scientist  and chairman of the collaboration board of the MAGIC telescopes.}

\vskip 1cm
\noindent
{\em \footnotesize The contributions of Giovanni Busetto, Mos\'e Mariotti, Gigi Peruzzo, Giorgio Sartori, Angelo Scribano, Riccardo Paoletti, and in particular of Luisa Bonolis, are acknowledged. Fig. 1 and 2 are from Marcello Cresti; Fig. 3 comes from the archives of the Padua University.}
\end{quotation}

\end{document}